\documentclass[10pt]{article}
\usepackage{amsmath}
\usepackage{graphicx}
\numberwithin{equation}{section}
\begin{document}
\title{\bf Lotka-Volterra dynamics under periodic influence}
\author{Debabrata Dutta$^1$ and J K Bhattacharjee$^2$\\\\$^1$S.N. Bose National Centre for Basic Sciences,\\ Saltlake, Kolkata 700098, India: debabrata@bose.res.in\\$^2$Indian Association for the Cultivation of sciences\\ Kolkata 700032, India: tpjkb@mahendra.iacs.res.in}
\maketitle
\begin{abstract}
Lotka Volterra model and its modified forms have long become a major area of interest for periodic motions in nonlinear systems with competitive species. The model given by Volterra shows that its periodicity is dependent on initial condition. This characteristics allows us to calculate the effect of periodic seasonal changes on population densities of different species.  
\end{abstract}
\section{Introduction}
In the third decade of the last century Lotka \cite{bib1} and Volterra \cite{bib2} formulated a coupled set of equations to describe an auto catalytic model and the statistics of fish catches in Adriatic. Since then Lotka Volterra model has become a central area of interest for periodic oscillations in nonlinear systems with competitive elements \cite{bib3}. It has found its applications in population biology \cite{bib4}, ecology \cite{bib5,bib6}, mathematical biology \cite{bib7}. Nevertheless It has often been criticized for being biologically unrealistic and mathematically unstable. Plenty of modifications were made to make it more realistic. Yet in last two decades area of application of Lotka-Volterra model expanded. It has made its way through newly explored applications, from membrane dynamics of competing neurons \cite{bib8}, neural networks \cite{bib9} , metabolic algorithm \cite{bib10} to network-electronics \cite{bib11,bib12} and stochastic dynamics \cite{bib13}.
    With this re-emergence  of Lotka-Volterra dynamics, we study some unexplored aspect of the original model.\\
Lotka-Volterra dynamics describes a predator-prey system where the prey thrives on the naturally available resources while predator thrives only on its interaction with the prey and would be extinct in a non interacting system. It is easy to realize that in such a system the prey population would thrive of the predators decrease while an increase in the predator number would cause the prey population to diminish. The predator  cannot flourish for ever because a decreasing number of prey would lead to fewer interaction and would drive the predator towards extinction. This produces the limit cycle like oscillations. The Lotka Volterra limit cycle is however very different from the usual limit cycles in that it depends on the initial conditions. This makes Lotka Volterra dynamics interesting. The unbounded growth of the prey population in the absence of the predator depends on a constant source of nutrients. In any practical situation, there will be some periodic fluctuations in the nutrients because of natural causes. If it is a wolf-rabbit situation, then the rabbit's supply of grass will have a seasonal variation. Interestingly enough, the periodic variation in the growth rate of the prey has not been studied in literature. There has been study of random variation \cite{rand} but not periodic ones. In this work, we study the effect of periodic variations.
  In carrying out this analysis, we noticed that the standard techniques that are used in dealing with nonlinear oscillators (Poincare Lindstedt method \cite{bib14}, equivalent linearization \cite{bib15} have not been carried out for the Lotka volterra model. Accordingly in section 2. We apply the Lindstedt Poincare technique on the Lotka Volterra model and come up with an initial condition dependent oscillation period. Our analytic and numerical results agree. In section 3, we use the analytic tools of section 2 to study the dynamics of the modulated system and also carry out a numerical analysis to test our predictions. Since theunmodulated system shows an initial condition dependent limit cycle, the final state of the driven system is found to be initial condition dependent. In section 4 , we apply another specified technique which work for high frequency perturbation and show how the calculation numerics compare favourably. A brief conclusion is presented in section 5.

\section{Depencence of Initial Condition}
\begin{figure}[h]
\centering
\includegraphics[scale=.3,angle=270]{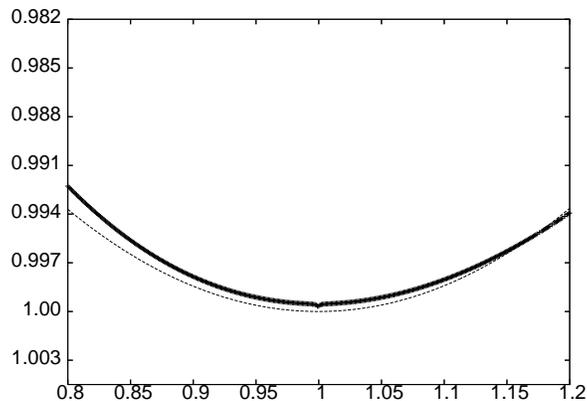}
\caption{Initial number of population density modifies its periodicity. Initial potulation density (x=y) is plotted with corresponding frequency ($\omega$) of its oscillation.}
\label{fig1}
\end{figure}

In this section, we focus on the limit cycle of the Lotka- Volterra model. As is well known, this limit cycle is unusual in the sense that it depends on the initial conditions. We show how the Poincare-linstedt technique can be applied to the model to obtain an initial condition dependent frequency for the limit cycle. This calculation is valid for small amplitudes and our numerical calculation supports the validity of this result.\\
 We take the simplest form of two species Lotka-Volterra system;
\begin{eqnarray}\label{eq1}
\begin{split}
\dot{x}&=x-xy\\
\dot{y}&=-y+xy
\end{split}
\end{eqnarray}
According to a linear stability analysis, the fixed point at (1,1) is a center. We first transform Eq(\ref{eq1}) to the variables $x_1=x-1$ and $y_1=y-1$ to write,
\begin{eqnarray}\label{eq2}
\begin{split}
\dot{x_1}&=-y_1-x_1y_1\\
\dot{y_1}&=x_1+x_1y_1
\end{split}
\end{eqnarray}
We note immediately that if we drop the nonlinear terms, then we have a simple harmonic oscillator of frequency unity. We are now in a position to carry out a Poincare-linstedt analysis. We imagine the existence of a parameter $\lambda$ multiplying the non-linear terms in Eq(\ref{eq3}) and introducing the frequency $\omega$ of the full dynamics, rewrite Eq(\ref{eq2})in the form;
\begin{equation}\label{eq3}
L\left( \begin{array}{c}x_1\\y_1\end{array} \right)=
\left( \begin{array}{ccc}
\frac{\partial}{\partial t} & \omega \\
-\omega & \frac{\partial}{\partial t}  
\end{array} \right)\left(\begin{array}{c}
x_1\\y_1\end{array} \right)= \lambda\left(\begin{array}{c}-1\\1\end{array} \right)x_1y_1+(\omega-1)\left(\begin{array}{c}y_1\\-x_1\end{array} \right)
\end{equation} 
For $\lambda\ll$1 we expand
\begin{eqnarray}\label{eq4}
\begin{split}
\omega&=1+\lambda \omega_1+ \lambda^2 \omega2+\dots \\
x_1&=x_{10}+\lambda x_{11}+ \lambda^2 x_{12}+\dots\\
y_1&=y_{10}+\lambda y_{11}+ \lambda^2 y_{12}+\dots
\end{split}
\end{eqnarray}
The right hand side of Eq(\ref{eq4}) has term of order $\lambda$ and higher. At order unity, the solution is $x_{10}=ReAe^{i \omega t}$ ,$y_{10}=Re\frac{A}{i}e^{i \omega t}$, where A is the amplitude of motion. At $\mathcal{O}(\lambda$), we have 
\begin{equation}\label{eq5}
L\left( \begin{array}{c}x_{11}\\y_{11}\end{array} \right)
=\left( \begin{array}{c}-1\\1\end{array}\right)x_{10}y_{10}+\omega_1\left(\begin{array}{c}
y_{10}\\-x_{10}\end{array} \right)
\end{equation} 
The solvability condition for an inhomogeneous second order differential equation now leads to $\omega_1=0$. The solution for $x_{11}$ and $y_{11}$ is found to be 
\begin{eqnarray}\label{eq6}
\begin{split}
x_{11}=Re\frac{A^2}{12i\omega}(1+2i)e^{2i\omega t}\\
y_{11}=Re\frac{A^2}{12i\omega}(1-2i)e^{2i\omega t}
\end{split}
\end{eqnarray}
At$\mathcal{O}(\lambda^2)$
\begin{equation}\label{eq7}
L\left( \begin{array}{c}x_{12}\\y_{12}\end{array} \right)
=\left( \begin{array}{c}-1\\1\end{array}\right)(x_{10}y_{11}+x_{11}y_{10})+\omega_2\left(\begin{array}{c}
y_{10}\\-x_{10}\end{array} \right)
\end{equation}
The solvability condition requires that the $e^{i\omega t}$ part of the right hand side be orthogonal to the left eigenvector of L. This leads to
\begin{equation}\label{eq8}
\omega_2=-\frac{|A|^2}{12\omega}
\end{equation}
The perturbative result for $\omega$ up to $\mathcal{O}(\lambda^2)$, after setting $\lambda=1$ is
\begin{equation}\label{eq9}
\omega=1-\frac{|A|^2}{12\omega}
\end{equation}
where A is the amplitude of the limit cycle. If $x_0$ and $y_0$ be the initial values of x and y, then we can write
\begin{equation}\label{eq10}
\omega=1-\frac{(x_{0}-1)^{2}+(y_{0}-1)^{2}}{12}
\end{equation}
We have checked this result numerically. The result are shown in Fig[\ref{fig1}]. The good agreement between the computed frequencies and the obtained from Eq(\ref{eq10}) is apparent.
\section{The modulated system at low and moderate frequencies}
With the help of the amplitude dependent frequency of the previous section, we explored Lotka-Volterra population dynamics under external drive. These systems have been extensively studied under effect of random perturbation \cite{rand} whereas under periodic forcing they were studied much less \cite{period}. We  explored it in presence of periodic perturbation. Periodic perturbation leads to diurnal or annual influences on the predator-prey systems and leads to periodic intrinsic growth rate in the prey population. \\
Under periodic forcing, the system [Eq(\ref{eq1})] becomes
\begin{eqnarray}\label{eq31}
\begin{split}
\dot{x}&=x-xy+\epsilon x\cos{\Omega t}\\
\dot{y}&=-y+xy
\end{split}
\end{eqnarray} 
Shifting to the ($x_1,y_1$) variables, we have
\begin{eqnarray}\label{eq32}
\begin{split}
\dot{x_1}&=-y_1-x_1y_1+\epsilon(1+x_1)\cos{\Omega t}\\
\dot{y_1}&=x_1+x_1y_1
\end{split}
\end{eqnarray} 
We note that if $\epsilon=0$ then the solutions can be written as periodic trajectories around (1,1) with a frequency that is dependent on the initial conditions. For $\epsilon=0$ we have equivalent oscillator 
\begin{eqnarray}\label{eq33}
\begin{split}
\dot{x_1}&=-\omega y_1\\
\dot{y_1}&=\omega x_1
\end{split}
\end{eqnarray}
with $\omega$ given by Eq(\ref{eq10}). In this technique of equivalent linearisation Eq(\ref{eq33}) can be rewritten as;
\begin{eqnarray}\label{eq34}
\begin{split}
\dot{x_1}&=-\omega y_1+\epsilon(1+x_1)\cos{\Omega t}\\
\dot{y_1}&=\omega x_1
\end{split}
\end{eqnarray}
Eliminating $y_1$ we have
\begin{equation}\label{eq35}
\ddot{x_1}-\epsilon\cos(\Omega t) \dot{x_1}+(\omega^2+\epsilon\Omega\sin{\Omega t})x_1=-\epsilon \Omega{\sin\Omega t}
\end{equation}
Defining $x_1=z-\frac{\epsilon\Omega\sin{\Omega t}}{\omega^2-\Omega^2}$ then,
\begin{equation}\label{eq36}
\ddot{z}+\omega^2z-\epsilon\frac{\partial}{\partial t}(z\cos{\Omega t}-\frac{\epsilon\Omega\sin{2\Omega t}}{2(\omega^2-\Omega^2)})=0
\end{equation}
When the driving amplitude is small,($\epsilon^2 \ll \epsilon$)
\begin{equation}\label{eq37}
\ddot{z}-\epsilon\cos\Omega t\dot{z}+(\omega^2+\epsilon\Omega\sin\Omega t)z=0
\end{equation}
Using the transformation $z=e^{\frac{\epsilon}{2\Omega}\sin{\Omega t}}p(t)$, we arrive at
\begin{equation}\label{eq38}
\ddot{p}+(\omega^2+\epsilon\Omega\sin{\Omega t})p=0
\end{equation}
correct to $\mathcal{O}$($\epsilon$). We recognize Eq(\ref{eq38}) as a Mathieu equation and note that there will be a periodic response at a frequency of $\frac{\Omega}{2}$ provided we fulfill the condition
\begin{equation}\label{eq39}
\begin{split}
\omega=\frac{\Omega}{2}(1\pm \epsilon)\text{\hspace{2cm}   or}\\
1-\frac{(x_{0}-1)^{2}+(y_{0}-1)^{2}}{12}=\frac{\Omega}{2}(1\pm \epsilon)
\end{split}
\end{equation}
This is in effect a construct on the initial conditions. The result found in Eq(\ref{eq39}) implies that for every modulating frequency $\Omega$, there will be some initial conditions for which a periodic response will be possible. For initial conditions in the range $\frac{\Omega}{2}(1-\epsilon) < 1-\frac{(x_{0}-1)^{2}+(y_{0}-1)^{2}}{12} < \frac{\Omega}{2}(1+\epsilon)$, the response will be unbounded while for initial conditions outside this range the response will be bounded and in general quasi-periodic - the periodic motion resulting when Eq(\ref{eq39}) is satisfied.\\
In Fig[\ref{fig2}],  the unboundedness of the solution is shown for $\Omega=2$. The width of the unbounded region is exactly $2\epsilon$. The original set of equation have nonlinear terms that prevents the dynamics becoming unbounded. Nevertheless we get the signature of unboundedness through the steep increase of the width ($\eta$) Fig[\ref{fig3}] of the phase space trajectory in the same frequency range.
\begin{figure}[h]
\centering
\includegraphics[scale=.3,angle=0]{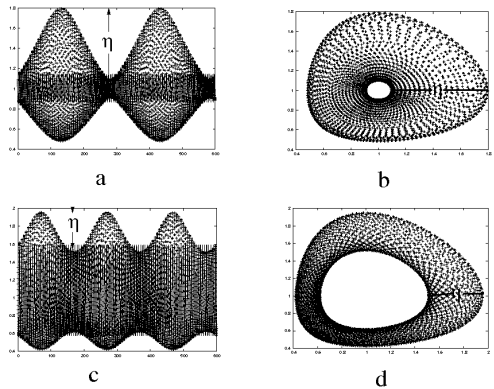}
\caption{Time series (a) of prey population ($x_1$) shows the signature of unboundedness of the linearized solution.Fig(\ref{fig3}) Change of width ($\eta$) (a,b) reflects the resonance in linearized rezime. As we move from linearized unbounded solutions $\eta$ decreases (c,d). solid lines (a,c) represent the unperturbed trajctories.}
\end{figure}
\label{fig2}
\begin{figure}[h]
\centering
\includegraphics[scale=.3,angle=0]{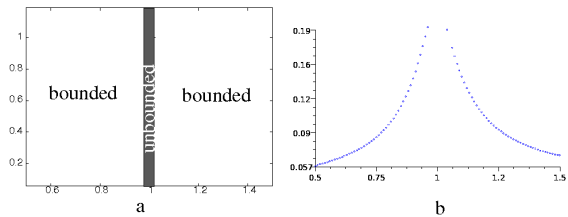}
\caption{a) Bounded  and unbounded solution from Mathieu equation [Eq(\ref{38})] campares well with the signature of resonance in  original dynamics (b). In (b) $\eta$ is plotted with initial conditions of population density ($x_0=y_0$).}
\label{fig3}
\end{figure}
$\eta$ is potentially very different from the width $\xi$ we discussed later. $\eta$ is a effect of the dynamics of two competing frequencies having resonating effect at comparable values, whereas $\xi$ shows the presence of two different order of timescale.
\section{High Frequency Limit}
\begin{figure}[h]
\centering
\includegraphics[scale=.3,angle=0]{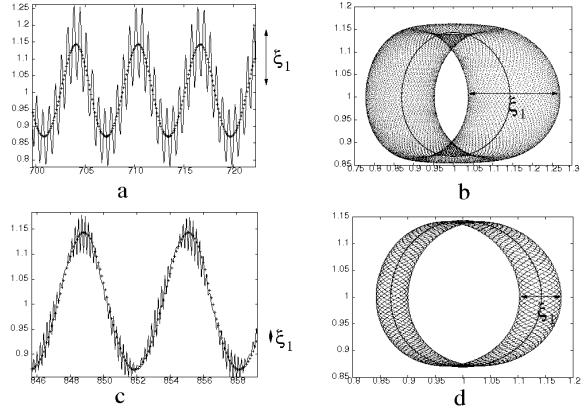}
\caption{(a,c)Time series shows how the amplitude ($\xi_1$) of high frequency perturbation changes as we vary frequency of perturbation ($\Omega$) from 10 (a) to 30 (b). Phase space plots (b,d) also show these through its change of width. the solid lines represent the unperturbed trajectory. }
\label{fig4}
\end{figure}

\begin{figure}[h]
\centering
\includegraphics[scale=.3,angle=270]{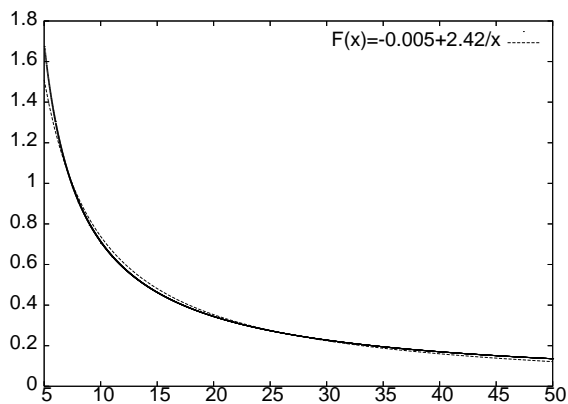}
\caption{ width $\xi_1$ is plotted against the frequency of modulation ($\Omega$). The fitting function (F(x)) supports the form of analytical result.}
\label{fig5}
\end{figure}

In this section, we consider the variation in the parameter of the Lotka Volterra model to be very rapid, i.e. the period $\Omega$ of the forcing in Eq(\ref{eq41}), is much greater than the frequency of the unforced system. The system fluctuates rapidly from the unperturbed trajectory and the phase space trajectory is broadened Fig[\ref{fig4}]. This broadening ($\xi$) depends mainly on the frequency of forcing and feebly on forcing amplitude. As $\Omega$ increases this broadening decreases very rapidly. 
The analysis of the system follows a procedure explained by Landau and lifshitz. We split the variables $x_1$ and $y_1$ into two parts,
\begin{eqnarray}\label{eq41}
\begin{split}	
x_1=X_1+\xi_1\\
y_1=Y_1+\xi_2
\end{split}
\end{eqnarray}
where  $\xi_1$, $\xi_2$ carry the rapid variations (scale of  $\Omega$) and the averaged quantities $X_1$ and $Y_1$ carry the slow variation (order unity) of the original model. Substituting in Eq(\ref{eq41});
\begin{eqnarray}
\dot{X_1}+\dot{\xi_1}&=&-Y_1-\xi_2-(X_1+\xi_1)(Y_1+\xi_2)+\epsilon(1+X_1+\xi_1)\cos{\Omega t}\label{eq42}\\
\dot{Y_1}+\dot{\xi_2}&=&X_1+\xi_1+(X_1+\xi_1)(Y_1+\xi_2)\label{eq43}
\end{eqnarray}
In this section, we do not restrict $\epsilon$ to be small. Instead, we note that since $\xi\sim\cos{\Omega t}$, typically $\dot{\xi}\gg \xi$. We choose the dynamics of $\xi_{1,2}$ to be
\begin{eqnarray}
\dot{\xi_1}&=&\epsilon X_1\cos{\Omega t}-X_1\xi_2\label{eq44}\\
\dot{\xi_2}&=&Y_1\xi_1\label{eq45}
\end{eqnarray}
In the above $X_1$ and $Y_1$ may be treated as constants on the scale of variation over a period $\Omega^{-1}$. In that situation, Eqs(\ref{eq44}) and (\ref{eq45}) reduce to 
\begin{equation}
\ddot{\xi_1}+Y_1 X_1\xi_1=-\epsilon X_1\sin{\Omega t}\label{eq46} 
\end{equation}
Since $\Omega\gg\sqrt{X_1Y_1}$, we write the approximate solution of Eq(\ref{eq46}) as
\begin{equation}
\xi_1=\frac{\epsilon X_1}{\Omega}\sin{\Omega t}\label{eq47}
\end{equation}
We now average Eqs(\ref{eq42}) and (\ref{eq43}) over the fast variations, drop terms that are $\mathcal{O}(\Omega^{-2})$ and obtain
\begin{eqnarray}\label{eq48}
\begin{split}
\dot{X_1}=-Y_1-X_1Y_1\\
\dot{Y_1}=X_1+X_1Y_1
\end{split}
\end{eqnarray}
which is the unforced system in the coarse-grained variables. The picture, which emerges is as follows: under a high frequency modulation, the motion can be explained as a rapidly oscillatory motion which is virtually identical to the trajectory of the unforced system. The amplitude of the fast variations decays according to $\frac{1}{\Omega}$ as shown in Eq(\ref{eq47}). Numerical simulation of the system bears out these expectations.Fig[\ref{fig5}]  
\section{conclusion}
In this paper we have shown that under lotka-volterra dynamics populations of predator and prey depend on their initial population densities. This dependence of initial population brings forth changes in dynamics but helps us formulate an effective Mathew-Hill type equation for this 2-species system under periodic seasonal changes. This also indicates slow seasonal changes have more effects in the variation of population than its faster counterpart.

\end{document}